# Privacy Protection of Automotive Location Data Based on Format-Preserving Encryption of Geographical Coordinates

Haojie Ji, Long Jin, Haowen Li, *Member*, *IEEE*, Chongshi Xin, and Te Hu

*Abstract*—There are increasing risks of privacy disclosure when sharing the automotive location data in particular functions such as route navigation, driving monitoring and vehicle scheduling. These risks could lead to the attacks including user behavior recognition, sensitive location inference and trajectory reconstruction. In order to mitigate the data security risk caused by the automotive location sharing, this paper proposes a high-precision privacy protection mechanism based on format-preserving encryption (FPE) of geographical coordinates. The automotive coordinate data key mapping mechanism is designed to reduce to the accuracy loss of the geographical location data caused by the repeated encryption and decryption. The experimental results demonstrate that the average relative distance retention rate (RDR) reached 0.0844, and the number of hotspots in the critical area decreased by 98.9% after encryption. To evaluate the accuracy loss of the proposed encryption algorithm on automotive geographical location data, this paper presents the experimental analysis of decryption accuracy, and the result indicates that the decrypted coordinate data achieves a restoration accuracy of 100%. This work presents a high-precision privacy protection method for automotive location data, thereby providing an efficient data security solution for the sensitive data sharing in autonomous driving.

This work was supported in part by the National Natural Science Foundation of China under Grant 52102447, in part by the R&D Program of Beijing Municipal Education Commission under Grant KM202411232005, and in part by the 2023 Industrial Internet Enterprise Data Security Joint Management and Control System Project under Grant 0747-2361SCCZA195. (Corresponding author: Long Jin.)

Haojie Ji is with the Key Laboratory of Modern Measurement & Control Technology, Ministry of Education, Beijing Information Science & Technology University, Beijing 100192, China (e-mail: jihaojie@bistu.edu.cn).
Long Jin is with the Key Laboratory of Modern Measurement & Control Technology, Ministry of Education, Beijing Information Science & Technology University, Beijing 100192, China (e-mail: 2023020008@bistu.edu.cn).
Haowen Li is with the the College of Faculty of Engineering, Hong Kong Polytechnic University, Hong Kong 999077, China (e-mail: howhen.li@connect.polyu.hk).
Chongshi Xin is with the Key Laboratory of Modern Measurement & Control Technology, Ministry of Education, Beijing Information Science & Technology University, Beijing 100192, China (e-mail: 2023020128@bistu.edu.cn).
Te Hu is with the Key Laboratory of Modern Measurement & Control Technology, Ministry of Education, Beijing Information Science & Technology University, Beijing 100192, China (e-mail: 2023020004@bistu.edu.cn).

## I. INTRODUCTION

AUTONOMOUS vehicles make it possible to collect and process the automotive location data accurately and in real time when the technologies such as automotive sensor networks (ASN), inertial navigation systems (INS), and high-precision global positioning systems (GPS) work together. This robust location data supports the critical applications including fleet management, real-time navigation, traffic flow prediction, and path planning. However, the privacy and security of users are exposed to substantial risks during the continuous and high-frequency collection and transmission of location data [1][2]. The sensitive information, such as a user's place of employment, social network, and even future whereabouts, could be got by attackers by analyzing an automobile movement trajectory, which can provide insight into the driving habits and lifestyle patterns of drivers. Therefore, autonomous driving is facing with the security challenge of vehicle sensitive data, which will bring serious threats to the personal privacy and driving safety of users [3][4].

Currently, there are four techniques usually used to preserve the privacy of location data, such as data anonymization, data obfuscation, differential privacy, and data encryption. Data anonymization obscures personal characteristics by grouping individuals together [5][6]. However, when applied to continuous vehicle trajectories, this method may greatly reduce the utilization efficiency of data while needing to continuously improve the intensity of privacy protection. By using regional generalization or adding noise, data obfuscation could obscure the actual location of vehicles [7][8]. But the availability of data becomes poor when meeting the requirement of privacy protection regulations. Differential privacy frequently creates excessive noise when applied to location data, impeding features like path planning and high-precision navigation [9][10]. Data encryption could offer excellent privacy protection. But this method will generally modify the specific format of encrypted data [11][12], thus degrading the availability of the decrypted data.

Generally, the format preservation of encrypted location data is crucial for the availability of private data in the multimodal data application of autonomous driving. Preserving format consistency is necessary for the inherent range constraints of geographical coordinates. If the format consistency of the encrypted location data is disturbed, the vehicle trajectory map projections may become distorted.

Format-preserving encryption (FPE) satisfies the requirement of data format consistency, but it still faces challenges when applied to automotive location data [13][14]. For example, FF1 and FF3-1 are two typical FPE algorithms that were initially developed for discrete finite sets [15]. However, the encryption of continuous floating-point data in latitude and longitude for FPE is difficult to effectively solve. There will be irreversible truncation errors due to a fixed scale factor when the preprocessing steps involve floating-point to integer conversion before encryption. Furthermore, coordinate errors may occur when values fall outside acceptable longitude and latitude ranges. This problem arises from the absence of encrypted data domain mapping methods. These methods are necessary to handle the distance limitations of geographical coordinate systems. The loss of accuracy in decryption is mainly caused by two factors. First, truncation errors may occur during data preprocessing. These errors have a growing effect on the decryption process. The effect grows in a nonlinear way. Second, floating point numbers have limits in binary encoding. This causes errors in the least significant bits. Both factors create meter-level position errors. This level of error is too large for vehicle navigation. High-precision positioning systems need much better accuracy.

In order to address the accuracy fidelity and format preservation consistency of automotive location data, this paper proposes a high-precision FPE designed specifically for latitude and longitude data. This method could maximize data format integrity and numerical accuracy while protecting location privacy at the same time. The primary contributions of this work are summarized as follows:

1) This paper proposes a geographic coordinate FPE method for automotive location data. The encrypted value is connected with the coordinate ID. This builds a bi-directional unique mapping relationship. This significantly solves the difficulty in FPE of accuracy loss when encrypting floating-point data. This method uses a thread-safe mapping mechanism. It provides accurate reversibility of the location encryption. This ensures that the decrypted values of latitude and longitude match the original data.
2) This paper presents a smart constraint method for geographic range of vehicle locations. The encryption parameters are dynamically modified for range intervals of varying longitudes and latitudes. This ensures that the encrypted data strictly adhere to the valid range of geographical coordinates. This method can ensure the validity of geographic data without reducing the encryption intensity. It uses the distance modulus operation in the last transformation stage of the Feistel network.
3) This paper provides an encryption optimization method to improve the privacy security. It applies the method to the round key selection and dynamic shift operations. The experimental results show that the proposed method improves encryption intensity while reducing processing complexity at the same time. In addition, this method has been proved to successfully solve the relationship between the spatial and clustering characteristics of the original location data. It achieves high-intensity privacy protection while maintaining the format and decryption accuracy of the location data.

The structure of this paper is as follows. Section I outlines the significance of FPE for automotive location data and investigates the challenges in existing methods. Section II reviews the relevant literature on location privacy protection and FPE, highlighting the persistent gaps in this research area. Section III elaborates on the design and implementation of the proposed FPE algorithm based on geographical boundary constraints. Section IV conducts the experiments and verifies the performance of the proposed algorithm. Finally, Section V presents the conclusion and future work.

## II. RELATED WORK

The research work on privacy protection of locations based on encryption methods has received increasing attention after the introduction of data security regulations. The methods such as K-anonymity, differential privacy, and spatial obfuscation could achieve privacy protection of location information through generalization or noise injection [16][17]. However, these methods generally compromise the accuracy of data or disrupt the characteristics of data formats. So it is difficult to meet the millimeter-level positioning accuracy and real-time positioning requirements in autonomous driving. Since the shared location information needs to retain a specific data format during encryption, FPE is regarded as a potential solution. Nevertheless, the FPE algorithm is mainly designed for discrete values, and it is difficult to adapt to the continuity of geographical coordinates for the conversion from floating-point to integer types. Regarding this issue in FPE, there are some schemes proposed by dynamic scaling factors or post-processing correction, but still cannot resolve the contradiction between the encryption strength of location data and the availability of privacy information.

*A. Data Anonymization*

The K-anonymity method hides the true location of specific vehicles. It generalizes the position to an area shared by at least K vehicles. This makes it difficult for attackers to identify the target vehicle [18]. Huang et al. [19] proposed a privacy-preserving trajectory synthesis method, and constructed trusted trajectories by extracting semantic, geographic, and directional similarities from real trajectories. Kuang et al. [20] introduced a location semantic privacy protection model based on spatial influence (LSPPM-SI) to mitigate semantic attacks. The model has been proven to protect location semantic privacy effectively. It maintains the availability of traffic data at the same time. Liu et al. [21] presented a fully virtual K-anonymity algorithm. This algorithm constructs an anonymous region through positional offsets. It gets entropy values and tracking success rates comparable to the optimal K-anonymity algorithm. The method avoids communication costs. Although the K-anonymization method shows computational lightness and low communication overhead, it introduces considerable positioning inaccuracy. This happens by producing large generalized areas. Furthermore, attackers may exploit prior knowledge to gradually reduce the search range. This occurs in cases of overlapping regions or consecutive queries. This compromises the effectiveness of K-anonymity.

In addition, the spatial occlusion method can achieve data anonymization. It replaces the exact vehicle position with a broader region. This region encompasses the precise location. This increases the uncertainty of the position information. This region typically includes at least K points. It can also meet the predefined minimum area. This achieves a level of protection comparable to K-anonymity. Asassfeh et al. [22] proposed an effective spatial masking algorithm to protect location privacy. Nevertheless, the practical deployment of spatial masking in dynamic driving scenarios remains constrained by excessive computational demands and communication costs. While avoiding complex encryption, these methods inevitably increase additional overhead due to the significant reduction in positioning accuracy.

*B. Data Obfuscation*

The location data protection method based on obfuscation could ensure the confidentiality of location information by concealing the actual user location with a false one. Gao et al. [23] proposed a differential location privacy-preserving mechanism based on trajectory obfuscation (LPMT). This mechanism hides each intermediate point into a target fuzzy sub-region and uses the sliding window algorithm to extract intermediate points as trajectory features. Subsequently, Laplace sampling is done within the target fuzzy sub-region to generate obfuscated GPS data. Compared with the baseline mechanisms, LPMT can reduce data quality loss by over 20% while maintaining the same level of obfuscation quality. Benarous et al. [24] introduced a cloud-enabled internet of vehicles (CE-IoV) location privacy protection solution that uses collaboration, obfuscation, and silence to minimize connectivity and tracking. Similarly, Zhao et al. [25] introduced a strict indistinguishability guarantee to enhance location data privacy while preserving high utility in directional distribution. This method used entry data obfuscation before dataset release and preserves the directional distribution with very high fidelity. However, location data privacy protection methods based on data obfuscation remain vulnerable to de-anonymization and attacks. This vulnerability exists due to the presence of persistent identifiable patterns or contextual associations.

*C. Differential Privacy*

Differential privacy limits the inference of a single true location by adding random noise to location queries or statistical results. The algorithm based on location clustering added Laplacian noise at the cluster center position and could protect the privacy of user trajectories while ensuring the quality of service [26]. Fang et al. [27] proposed a location data collection method based on local differential privacy. They verified the feasibility on three real-world public check-in datasets. The experimental results proved that the proposed method retained more than 90% of the data availability while protecting privacy. Similarly, Ju et al. [28] proposed an energy sharing scheme using local differential privacy (LDP). This scheme eliminated the need for a trusted third party and effectively resisted multiple types of attacks. These attacks included trajectory reasoning attacks, unified attacks, accidental attacks, traffic analysis-based attacks, auxiliary information attacks and data landmine attacks. Furthermore, this scheme used the privacy-weighted average model, reducing the distortion caused by LDP and improving the usefulness of trajectory data. However, while differential privacy provides strong privacy guarantees that effectively prevent attackers from tracking specific locations, the built-in noise injection mechanism always reduces positioning accuracy. Specifically, the buildup of noise in later locations may lead to a decrease in the available data. Furthermore, the distribution of privacy budgets determines the strength and effectiveness of privacy protection. The latency of location services will increase in autonomous driving with high real-time requirements. This increase is due to budget management and noise calculation.

*D. Data Encryption*

Homomorphic encryption [29] enables direct calculation on encrypted automotive location data without the need for decryption.

To address emerging security challenges including data breaches and quantum computing threats, Palma et al. [30] proposed a privacy protection framework. This framework guaranteed continuous data encryption during matching operations. Singh et al. [31] proposed a homomorphic encryption framework for GPS trajectory anomaly detection that preserved location privacy during the entire analysis process. Rupa et al. [32] proposed a homomorphic encryption method based on matrix transformation. In this method, each character in plain text had its binary converted ASCII value shifted, rotated, and transposed. In symmetric encryption, the same key is used for both encryption and decryption. Due to this method, an attacker cannot easily predict the plaintext through statistical analysis. The advantage of homomorphic encryption is its high security level while eliminating the need to trust a third party. However, its limitation lies in the extremely high computational complexity and significant delays caused by ciphertext operation and noise management. In vehicular systems, the high computing and communication overhead may become the bottleneck for the real-time performance and scalability of the system.

FPE allows the original format of the data to be retained after encryption. Bellare et al. [33] formally defined the FPE framework and its security objectives. Konduru et al. [34] proposed a privacy-preserving approach based on blockchain and FPE to protect users' data privacy. Yang et al. [35] evaluated the execution efficiency of the FF1 algorithm and identified the efficiency bottleneck. Based on the analysis results, optimization methods were given from the perspectives of presetting, algorithm structure, and format conversion function. Finally, the simulation results showed that the performance improvement after optimization was significant. The degree of performance improvement increased with the increase of the length of the plain text. Markani et al. [36] evaluated the feasibility of enhancing the automatic dependent surveillance-broadcast (ADS-B) security. They employed FPE, Feistel-based encryption, and multiple implementation difference algorithms. The provided solution has been implemented in the standard software-defined Radio ADS-B for real-time applications.

## III. Encryption Method

An encryption algorithm for automotive location data is introduced in this section. It is designed based on geographic range constraints. The proposed algorithm consists of three parts, i.e., the mapping mechanism based on composite keys, the constraint mechanism related to geographic range, and the Feistel network.

### A. Mapping Mechanism Based on Composite Keys

1) Design principle

FPE generally has the issue on decryption accuracy loss when dealing with data with decimal positions such as longitude and latitude of locations. FPE is able to construct a reversible mapping so that the data fields remain unchanged before and after encryption. However, the mapping may not be completely reversible. This may lead to differences between decrypted values and original values when applied to location data with high requirements for decimal accuracy. Therefore, this paper designs an exact mapping mechanism based on composite keys. The main idea of this mechanism is to construct an extended mapping function $M$ as follows:

$$M: (id_{coord}, v_{enc}) \rightarrow v_{orig}, \quad (1)$$

Where, $id_{coord}$ is the unique identifier of the coordinate, $v_{enc}$ and $v_{orig}$ are the encrypted value and the original value, respectively. This composite key is designed linking the encrypted value to the coordinate identifier, thereby creating an exact one-to-one mapping relationship.

To achieve precise mapping of each part of the longitude and latitude coordinates, this paper defines four independent mapping functions as follows:

$$M_{lon}^{int}: (id_{coord}, v_{enc}) \rightarrow v_{orig}^{lon\_int}, \quad (2)$$

$$M_{lon}^{frac}: (id_{coord}, v_{enc}) \rightarrow v_{orig}^{lon\_frac}, \quad (3)$$

$$M_{lat}^{int}: (id_{coord}, v_{enc}) \rightarrow v_{orig}^{lat\_int}, \quad (4)$$

$$M_{lat}^{frac}: (id_{coord}, v_{enc}) \rightarrow v_{orig}^{lat\_frac}, \quad (5)$$

In these equations, $M_{lon}^{int}$ and $M_{lon}^{frac}$ represent the integer and decimal parts of the longitude mapping function, respectively. $M_{lat}^{int}$ and $M_{lat}^{frac}$ represent the latitude mapping function's integer and decimal parts, respectively. In addition, $id_{coord}$ represents the unique identification of vehicle coordinates. $v_{enc}$ represents the encrypted value. $v_{orig}^{lon\_int}$ and $v_{orig}^{lon\_frac}$ denote the integer and decimal parts of the original longitude, respectively. $v_{orig}^{lat\_int}$ and $v_{orig}^{lat\_frac}$ denote the integer and decimal portions of the original latitude, respectively.

The composite key mapping algorithm receives information about the original value, encrypted value, coordinate ID, and coordinate type. This algorithm selects the appropriate mapping function based on the coordinate type and records the mapping relationship between the composite key and the original value. Meanwhile, the algorithm detects and records mapping conflicts, which occur when the same encryption value comes from different coordinates. Although there is a mapping conflict, the composite key is able to perform correct decryption.

2) Classified storage strategy

This paper proposes a classified storage strategy. The latitude and longitude data are broken down into decimal and integer components. Both components are processed independently. The following is a definition of the decomposition function $D$ for the vehicle position coordinate presented as $C$:

$$D(C) = (sign(C), int(|C|), frac(|C|), d), \quad (6)$$

Where, $sign(C)$ represents a symbolic function with values of 1 or -1, $int(|C|)$ represents the integer part after taking the absolute value, $frac(|C|)$ represents the decimal part after taking the absolute value, and $d$ represents the number of digits in the decimal part.

The original coordinates could be restored through the recombination function $R$ as follows:

$$R(sign, int, frac, d) = sign \cdot (int + frac \cdot 10^{-d}), \quad (7)$$

Where, $sign$ represents the symbol, $int$ represents the integer part, $d$ represents the decimal part.

The encryption process independently applies encryption functions $E$ to each decomposed part, which can be expressed as:

$$E_C = R(sign(C), E(int(|C|), k_1), E(frac(|C|), k_2), \quad (8)$$

Where, $E_C$ represents the coordinate encryption function, $E$ represents the basic encryption function, $k_1$ and $k_2$ represent the encryption parameters of the integer part and the decimal part respectively.

The coordinate encryption algorithm first deals with the sign of the position coordinates. Then it breaks down the coordinates into integer and fractional components. Each component is encrypted separately using the encryption function while maintaining the original format. The components are then combined into encrypted coordinates. To guarantee that the encrypted results are in the same format, the algorithm applies the appropriate range constraints and masks based on the type of coordinates and values during the encryption process.

3) Mapping conflict detection mechanisms

There may be mapping conflicts as a result of different original values producing the same encrypted value in encryption process. This paper designs two conflict handling mechanisms.

The first mechanism is composite key differentiation. This design can differentiate the encrypted values through the coordinate ID, even if they are the same. For two different coordinates presented as $id_1$ and $id_2$, they can still be distinguished through the composite key even if the encrypted values are the same. This mechanism can be presented as follows:

$$M(id_1, v_{enc}) = v_1 \land M(id_2, v_{enc}) = v_2, v_1 \neq v_2, \quad (9)$$

Where, $M$ represents the mapping function, $id_1$ and $id_2$ represent different coordinate ids, $v_{enc}$ represent the same encrypted value, $v_1$ and $v_2$ represent different original values.

The second mechanism is conflict statistics and monitoring. During the encryption process, the conflict status of various types of mappings in real-time is recorded. The definition of the conflict rate $CR$ is as follows:

$$CR = \frac{|\{v_{enc} | \exists i \neq j : v_{enc_i} = v_{enc_j}\}|}{|V_{enc}|}, \quad (10)$$

Where, $CR$ represents the conflict rate, $v_{enc}$ represents the encrypted value, $v_{enc_i}$ and $v_{enc_j}$ represent the $i$ and $j$ encrypted values respectively, and $V_{enc}$ represents the set of all encrypted values.

Conflict statistics include four types of mapping conflict counters $C_{lon}^{int}$, $C_{lon}^{frac}$, $C_{lat}^{int}$, and $C_{lat}^{frac}$, corresponding to the number of conflicts in the integer part of longitude, the decimal part of longitude, the integer part of latitude, and the decimal part of latitude, respectively. The conflict handling algorithm records conflict information when a mapping conflict is detected but does not interrupt the processing flow. During decryption, an exact composite key lookup is first used, and it may fall back to a fuzzy lookup if no match is found.

*B. Geographical Constraint Mechanism*

1) Longitude and latitude

The geographic coordinate system defines the strict numerical range of longitude and latitude. The FPE method may generate invalid coordinates beyond these ranges after encryption, resulting in an unsatisfactory encryption effect. In order to resolve this issue, this paper defines the following range sets based on the segmented distribution characteristics of the integer parts of longitude and latitude as follows:

$$R_1 = \{x \in Z | 0 \leq x < 10\}, \quad (11)$$

$$R_2 = \{x \in Z | 10 \leq x < 100\}, \quad (12)$$

$$R_3 = \{x \in Z | 100 \leq x < 180\}, \quad (13)$$

$$R_4 = \{x \in Z | 0 \leq x < 10\}, \quad (14)$$

$$R_5 = \{x \in Z | 10 \leq x < 90\}, \quad (15)$$

Where, $R_1$, $R_2$ and $R_3$ respectively represent the range of units, tens and hundreds of digits of longitude. $R_4$ and $R_5$ represent the single-digit and ten-digit ranges of latitude respectively.

$$RT(v, \lambda_{lon}, \lambda_{int}) = \begin{cases} 1, & if \lambda_{lon} \land \lambda_{int} \land v \in R_1 \\ 2, & if \lambda_{lon} \land \lambda_{int} \land v \in R_2 \\ 3, & if \lambda_{lon} \land \lambda_{int} \land v \in R_3 \\ 4, & if \neg \lambda_{lon} \land \lambda_{int} \land v \in R_4 \\ 5, & if \neg \lambda_{lon} \land \lambda_{int} \land v \in R_5 \\ 0, & otherwise \end{cases}, \quad (16)$$

Where, $RT$ represents the range type identification function, $v$ represents the value to be processed, and $\lambda_{lon}$ and $\lambda_{int}$ are boolean values indicating whether it is a longitude and an integer part, respectively. The function checks the geographic coordinate type and value range of the input value. It maps the value to the corresponding range type. This provides the foundation for subsequent range constraint processing. The range type determination algorithm first determines the coordinate type and value type and determines the specific range type based on the value. For values not within the specified range, the function returns the default type. This indicates that no special constraint handling is required. For example, when the input is the integer part of longitude and the value is 15, $\lambda_{lon} = true$, $\lambda_{int} = true$, and $15 \in R_2$, so $RT(15, true, true) = 2$.

2) Geographic coordinate

This paper proposes the geographical range constraint algorithm to improve encryption accuracy. The algorithm ensures that the encrypted coordinates remain within the valid range by combining modular arithmetic and radian offsets. The range constraint function is defined as follows:

$$RC(v', rt) = \begin{cases} v' \bmod 10, & if\ rt = 1 \\ 10 + (v' \bmod 90), & if\ rt = 2 \\ 100 + (v' \bmod 80), & if\ rt = 3 \\ v' \bmod 10, & if\ rt = 4 \\ 10 + (v' \bmod 80), & if\ rt = 5 \\ v', & if\ rt = 0 \end{cases}, \quad (17)$$

Where, $RC$ represents the range constraint function. $v'$ represents the encrypted intermediate value, which is the output after the Feistel network transformation. $rt$ represents the range type, that is, the output of the $RT$ function.

This constraint algorithm is mainly designed to map any input value to the specified valid range. This constraint algorithm has two main characteristics. First, this algorithm can maintain range validity. By using designed modulus values and base offsets, it can ensure that encryption results strictly follow the valid range of geographic coordinates. For example, the constraint algorithm ensures that its integer component does not exceed the maximum effective value of longitude or latitude. In addition, it can maintain consistent distribution. Modular arithmetic ensures that encryption results are roughly uniformly distributed within the effective range. This avoids excessive concentration or gaps in specific value ranges, thereby improving the randomness and security of encrypted values.

The range constraint algorithm is executed after the Feistel network transformation is completed. It is the last step of the entire encryption process. It ensures that the encrypted coordinates remain within the valid range of geographic coordinates. This addresses the issue that FPE may generate invalid values when processing geographic coordinates. This is crucial for ensuring that the encrypted data can be directly used for map rendering and spatial computing.

3) Dynamic Masking and Modular Arithmetic

To ensure the format consistency of encrypted data, this paper proposes a dynamic masking mechanism as a crucial complement to geographic range constraints. FPE algorithms are typically designed to preserve the data format via specialized substitution or transformation. However, the specific value range constraints of latitude and longitude should be considered when working with vehicle location data. The dynamic masking mechanism is proposed as an useful addition to the FPE algorithm. This method precisely controls the value range within the Feistel network structure. It guarantees that the encrypted data follows the specific format requirements for vehicle location.

The basic idea of dynamic mask selection method is to determine the precise binary mask bit width according to the numerical properties of the input data. The method uses a uniform 16-bit mask to represent the integer portion of the geographic coordinate. This is sufficient to cover both longitude and latitude maxima while ensuring a high level of security. For the fractional parts, the mask bit width is dynamically determined based on the value. An 8-bit mask is used for values less than 100. A 10-bit mask is used for values ranging from 100 to 1000. Therefore, a larger value requires a wider mask accordingly.

The mask is applied using a bitwise sum operation. This means that the intermediate value of the transformed Feistel network and the selected mask value are bitwise summed. This operation ensures that the resulting value remains within the expected bit-width range. This prevents potential bit overflow issues during the encryption process.

The dynamic mask design presented in this paper effectively complements the range constraint mechanism. The mask operation defines the upper bound of the bit width of the values and provides the foundation for the range constraints. Besides, the range constraint also accurately maps the values to valid intervals of geographic coordinates. This hierarchical constraint mechanism ensures that the encrypted data not only maintains consistent bit length, but also strictly adheres to the range requirements of geographic coordinates in terms of numerical values.

The dynamic masking mechanism is applied in each round of transformation within the Feistel network to ensure that the intermediate results remain within reasonable bounds. This not only improves the ability to control the data format, but also improves the algorithm's stability in real-world applications. In particular, the dynamic masking mechanism, demonstrates a significant advantage when dealing with information with a narrow numerical range, such as position data.

*C. Feistel Network Structure Improvement*

1) Calculation of tweak values

FPE algorithms typically employ a fixed encryption procedure and are difficult to account for the variety and correlation of automotive location data. To improve the security and anti-batch analysis capabilities of the FPE algorithm, a dynamic tweak generation and application mechanism based on coordinate characteristics is proposed.

The unidirectionality and avalanche effect of the hash function MD5 provide important guarantees. Even similar input values produce entirely different tweak values, which are the foundation of the tweak value calculation algorithm. First, the input value is concatenated with the key hash to generate a sequence of bytes. Then, the MD5 hash of the sequence is calculated, and the first four bytes of the hash value are extracted, and finally a 32-bit tweak integer is output.

In the encryption process, three synergistic mechanisms integrate the tweak values into the Feistel network structure. The initial hybrid transform servers as the first mechanism. In this case, the initial transformation of the entire network includes a direct XOR operation that combines the tweak values with the input values. This step increases the initial perturbation's unpredictability by disrupting the linear relationship between the input values. Dynamic key selection serves as the second mechanism. Input-dependent dynamic key scheduling is achieved by incorporating tweak values into the key selection process for every round. In particular, the tweak value and the round number work together to determine the key index used in each round. Even when the number of rounds stays constant but the input values change, different round keys can be used thanks to this method, thereby overcoming the limitation of fixed key sequences in conventional Feistel networks. The third mechanism involves the dynamic computation of the shift amount. The algorithm's diffusion properties are improved by using tweak values to determine the cyclic shift amounts in each operation cycle. By ensuring that even small changes in the input values spread more quickly across all of the output bits, this dynamic shift design greatly increases the resistance to differential analysis of the algorithm.

These three synergistic approaches work together to make the tweak value a key factor in the overall encryption process. Due to the generation of different tweak values, even the data from the same vehicle at similar positions in different times follow entirely distinct encryption paths. This effectively prevents attackers from employing batch analysis to identify encryption patterns.

2) Key selection and dynamic calculation of shift amount

The Feistel network usually employs a fixed key scheduling scheme. This results in different input data sharing the same key usage sequence, thereby reducing the encryption strength. In order to achieve input-adaptive key scheduling, we propose a dynamic key selection method. This method combines the current round number and adjustment value. Its main idea is to change the current round number by utilizing the selected bits of the adjustment value, and then determine the actual used round key index based on the pattern of the total number of rounds and keys.

This design makes use of the 32 rounds of keys of the SM4 algorithm and introduces a variable factor depending on the input data to disrupt its fixed order. For example, if the fine-tuning values extracted from the coordinates of two vehicles at different location are 0x12345678 and 0x87654321 respectively. Then, even in the same round, different round keys will be selected due to the difference in the lower 5 bits. This method significantly increases the unpredictability of the encryption process.

Similarly, the calculation of shift amount changes from a fixed value to a dynamic process. This method uses the lower 3 bits of the tweak value to tinker with the current number of rounds. Subsequently, the result is taken modeled against 7 and added to 1 to ensure that the final shift amount is between 1 and 7. This dynamic shift design significantly enhances the algorithm's diffusion properties and resistance to attack. For different input values, the shift amount may be different even at the same number of rounds. This will result in completely different displacement effects, thereby enhancing the security of the algorithm.

The dynamic calculation of key selection and shift amount working together can ensure that each round of transformation exhibits the uncertainty related to the input. This design can effectively enhance the security of the algorithm.

3) Optimization of the round function

This paper presents a comprehensive optimization design of the Feistel network's round function for vehicular encryption applications. The optimization results in a complete encryption structure that effectively adapts to the characteristics of vehicular location data. The optimized round function integrates the key XOR, dynamic circular shift, and mask application of the SM4 algorithm, thereby forming the format-preserving transformation.

The workflow of the round function consists of three steps. The first step involves performing an XOR operation between the input value and the round key selected for the current round. Since the key selection is influenced by the adjustment value, different round keys will be employed even for the same round number. The second step involves performing a circular left shift on the XOR result, where the shift amount is determined by the aforementioned dynamic shift amount calculation method. This dynamic shift not only enhances the data diffusion effect, but also accelerates the propagation of the avalanche effect. The third step involves applying a mask to the shifted result, thereby ensuring that the output value remains within the expected bit length. The mask value is determined according to the dynamic mask mechanism introduced earlier, thereby ensuring data format consistency.

These three steps work together to form a flexible transformation function. Specifically, the key XOR provides confusion, the circular shift provides diffusion, and the mask application ensures format preservation. This design fully leverages the advantages of cryptographic primitives while adapting to the special needs of positional data.

The complete encryption process consists of three stages, i.e., initial processing, multiple rounds of iteration, and final constraints. First, the input value is XORed with the adjustment value to produce the initial value. Then, the transformation is applied through multiple iterations of the wheel function. Finally, a range constraint function is applied to the final output to ensure the result falls within the valid range of geographical coordinates. The corresponding decryption process performs inverse operations in reverse order to restore the original data. The composite key mapping mechanism proposed in this paper effectively solves the irreversibility problem of range constraints.

The improved Feistel network proposed in this paper forms a more secure, flexible, and adaptable encryption structure for vehicular location data through value adjustment mechanisms, dynamic parameter calculations, and round function optimization. The adjustment value mechanism provides input-related random seeds for dynamic parameter calculations. These dynamic parameter calculations cause the round function to exhibit different characteristics in each round transformation. Subsequently, the optimized round function integrates these dynamic elements, thereby forming a complete encryption process.

IV. EXPERIMENTAL RESULTS AND ANALYSIS

*A. Experiment on Relative Distance Retention Rate (RDR)*

In order to effectively evaluate the protection level of spatial relations by the proposed trajectory encryption method, the relative distance retention rate (RDR) is introduced as an evaluation index in the experiments. RDR measures how well the encrypted trajectory maintains the relative distance relationships between point pairs compared to the original trajectory. This metric is of great significance for many location-based applications, which typically rely on the relative distance relationship between points rather than the absolute distance.

To ensure strong privacy protection, an optimal trajectory encryption method should significantly change the relative distance relationship between trajectory points while maintaining data availability. This method could effectively prevent attackers from inferring sensitive location information through relative distance analysis. More importantly, the geographic range constraint mechanism proposed in this study guarantees that each coordinate point after encryption remains within the valid range of the geographic coordinate system, thereby providing an additional protection for data security. This mechanism could effectively prevent attackers from making judgments whether these seemingly reasonable coordinates are encrypted or not when intercepting the data.

1) Experimental principle of RDR

The original trajectory $T_O = \{p_1, p_2, \cdots, p_n\}$ and the corresponding encrypted trajectory $T_e = \{p_1', p_2', \cdots, p_n'\}$ are given in this experiment. First, four different points $p_i$, $p_j$, $p_m$ and $p_n$ from the trajectory are randomly selected and written as two pairs of points $(p_i, p_j)$ and $(p_m, p_n)$.

TABLE I
DETAILED STATISTICAL RESULTS OF RDR VALUES

| Statistical indicators | Numerical value |
|---|---|
| Average value | 0.0844 |
| Upper quartile | 0.0515 |
| Standard deviation | 0.1071 |
| Minimum value | 0 |
| Maximum value | 1 |
| First quartile | 0 |
| Third quartile | 0.1324 |
| Number of vehicles with RDR of 0 | 3,360 |
| Ratio of RDR to 0 | 33.09% |
| Number of vehicles in service | 10,153 |

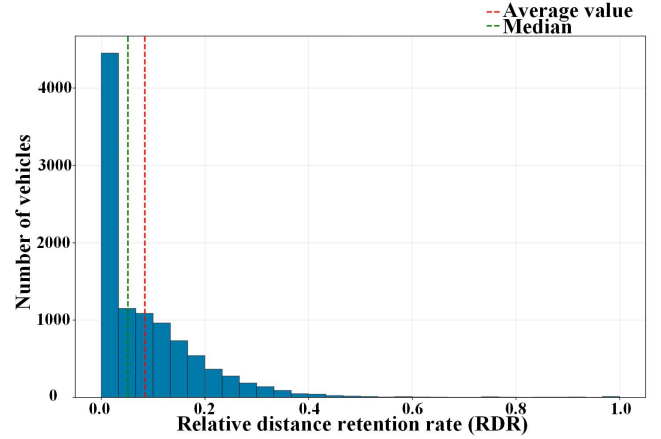

Fig. 1. Histogram of frequency distribution of RDR

In the original trajectory, the geographical distances $dist(p_i, p_j)$ and $dist(p_m, p_n)$ between each pair of points are calculated. The distance between two points on the Earth's surface is calculated by the Haversine formula as follows:

$$dist(p_i, p_j) = 2R \cdot arcsin\left(\sqrt{sin^2(\frac{\phi_j-\phi_i}{2}) + cos(\phi_i)cos(\phi_j)sin^2(\frac{\lambda_j-\lambda_i}{2})}\right), \quad (18)$$

Where, R represents the radius of the Earth. ϕ and λ represent the latitude and longitude of a point, respectively.

Then the distance ratios $r_o$ and $r_e$ of both two pairs of points are calculated in the original trajectory and the encrypted trajectory as follows:

$$r_o = \frac{dist(p_i,p_j)}{dist(p_m,p_n)}, \quad (19)$$

$$r_e = \frac{dist(p_1',p_2')}{dist(p_m',p_n')}, \quad (20)$$

The relative error $\epsilon$ can be represented by:

$$\epsilon = \frac{|r_o - r_e|}{r_o}, \quad (21)$$

After calculating the relative errors of multiple pairs of randomly sampled points, the average relative error is calculated, and finally the value RDR is represented by:

$$RDR = 1 - min(\bar{\epsilon}, 1), \quad (22)$$

2) Results of the experiment

The T-drive taxi trajectory dataset, including trajectory data from 10,308 vehicles, is used in this experiment. 10,153 valid results are obtained from the analysis of the trajectories of these 10,308 vehicles. The main statistical results of RDR values are shown in Table 1.

Experimental results show that the encryption method proposed in this paper achieves extremely low RDR values on most trajectories. The average RDR value is only 0.0844, indicating that the encryption algorithm effectively disrupts the relative distance relationship between the trajectory points and provides high-intensity privacy protection for the location data. It is particularly noteworthy that 33.09% of the trajectories have an RDR value of 0. This result indicates that the relative distance relationship of these trajectories is completely altered. And attackers could not obtain useful information by analyzing the distance relationship. The unique advantage of this algorithm is that each encrypted coordinate point is still a geographically valid GPS coordinate despite the spatial relationship being completely disrupted. This means that even if attackers who acquire the encrypted data may misinterpret it as real trajectory data from another region rather than recognizing its encrypted features. This camouflage greatly enhances the security of the data in the stages of transmission and storage.

Figure 1 illustrates the histogram of the frequency distribution of RDR, revealing the extent of the effect of the present encryption method on the spatial properties of the trajectories. The distribution shows obvious right-skewed characteristics, with a large number of samples concentrated in the region close to 0, and more than 65% of the samples have RDR values less than 0.1. This distribution pattern suggests that the relative distance relationship of most trajectories has changed significantly after encryption, which is highly consistent with the desired privacy protection effect. Meanwhile, a small number of trajectories exhibit RDR values exceeding 0.4, resulting in a long-tailed distribution characteristic. The right-skewed nature of the distribution is further confirmed by the fact that the mean value is significantly higher than the median. This suggests that a small number of trajectories with relatively high RDR values pull up the mean, while the median better represents the performance of typical trajectories.

The boxplot presented in Figure 2 reveals another perspective of the data distribution characteristics. The box is compressed in the low value region, further confirming the property that most trajectories have low RDR values. It is noteworthy that a large number of outliers are distributed in the high value region, which may represent special trajectory patterns or vehicles with unique spatial characteristics. The relatively small quartile differences indicate that the core data distribution is relatively concentrated and most trajectories exhibit similar low RDR characteristics.

As shown in Figure 3, the cumulative distribution function

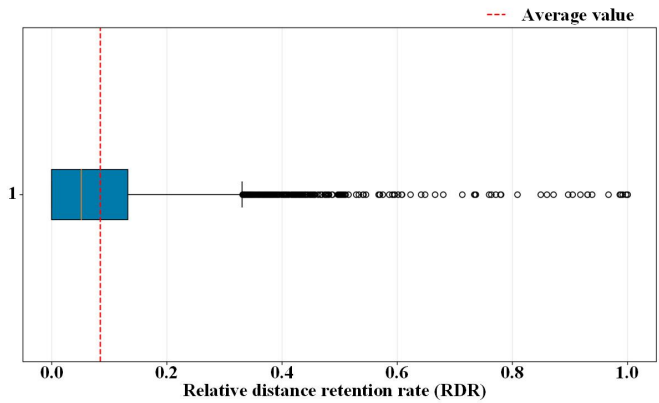

**Fig. 2.** Box plot of RDR value distribution

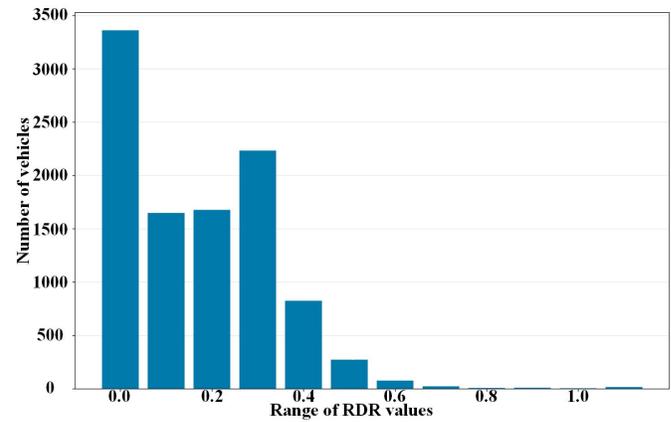

**Fig. 4.** Interval distribution statistics of RDR values

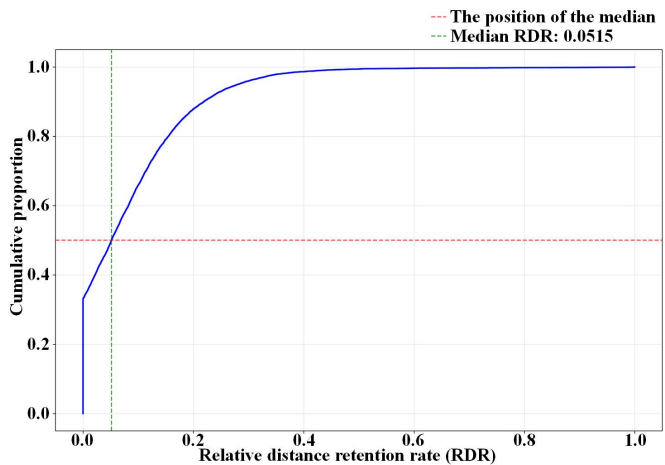

**Fig. 3.** CDF of RDR values

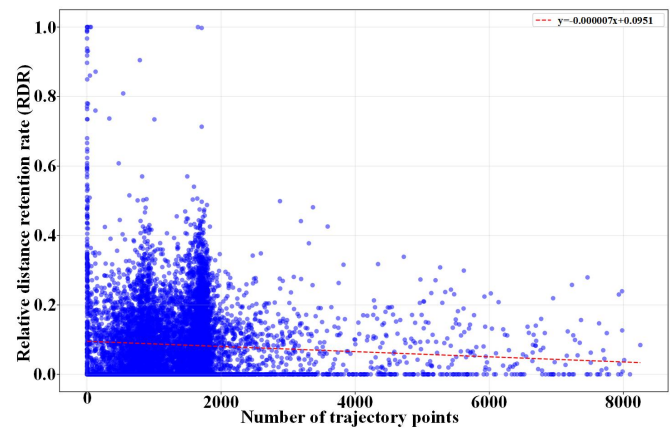

**Fig. 5.** Scatterplot of RDR values versus number of trajectory points

(CDF) of the RDR values yields a more accurate representation of the statistical distribution. With approximately 33% of the samples having an RDR value of 0, the CDF curve exhibits a distinct step pattern and is extremely steep in the region with low RDR values. The median RDR values across the samples is 0.0515, indicating that 50% of the samples have values of no more than this threshold. The fact that this median is significantly less than the optimal median value of 0.5, indicates that the proposed encryption method successfully messes with the spatial characteristics of the trajectories. The region of high RDR values is where the curve flattens down, indicating that samples with high RDR values are comparatively rare. According to the CDF curves, approximately 80% of the samples have RDR values not exceeding 0.2, 90% have RDR values not exceeding 0.3, and 95% have RDR values not exceeding 0.4. If the RDR value of 0.2 is considered the threshold for appropriate privacy protection, approximately 80% of trajectories meet the protection criterion. This indicates that the suggested strategy provides strong protection for the majority of trajectories.

As shown in Figure 4, the segmented statistics of RDR values display the number of samples in various RDR value intervals. There are 3,360 samples, 33.09% of the total amount, with an RDR of 0, indicating that the most frequent outcome is total disruption of the relative distance relationship. Approximately 8,100 samples have RDR values in the range of 0 to 0.2, accounting for roughly 80% of the total sample count. Specifically, there were approximately 2,330 samples in the 0 to 0.05 interval, around 1,220 samples in the 0.05 to 0.1 interval, and roughly 1,220 samples in the 0.1 to 0.2 interval. In contrast, there are approximately 230 samples, accounting for only 2.3%, with RDR values above 0.6. As the RDR value increases, the number of samples shows a significant exponential decline, indicating that the proposed encryption algorithm meets the requirements for robust privacy protection.

The scatterplot relationship between RDR values and the number of trajectory points is shown in Figure 5 to investigate any possible correlation between the two variables. The slope of the fitted trend line is slightly negative but close to zero. Although this association is weak, it indicates a minor negative tendency in RDR values as the number of trajectory points increases. In particular, the RDR value typically decreases by 0.07 for every additional 1,000 trajectory points, which represents a reduction that is essentially insignificant in practical applications. Besides, the data points exhibit extremely distributed characteristics, indicating that the RDR value is not determined by the trajectory's length. Additionally, the distribution of the trajectory length in the dataset is related to the point density, which is higher in the area with fewer trajectory points and decreases as the number of trajectory points increases. Given that longer trajectories present more complex and variable movement patterns, and encryption is more likely to break the initial relative distance connections, samples with more than 4,000 trajectory points typically have lower RDR values.

The results of RDR experiments strongly demonstrate the efficiency of the proposed location data encryption method in terms of privacy protection. The RDR experiments not only confirm the effectiveness of the algorithm in destroying spatial relationships, but also achieve a kind of stealth encryption protection by combining with the geographic range constraint mechanism. The encrypted data completely changes the semantic information while preserving the format validity. The combination of the two protection mechanism makes the algorithm have significant advantages in practical applications, e.g., the data encrypted by this mechanism could prevent privacy attacks based on spatial analysis, and also avoid exposing the encryption behavior due to the abnormal data formats.

*B. Experimental Design for Hotspot Area Identification and Analysis*

The hotspot area identification and analysis experiment evaluates the effectiveness of encryption algorithms considering the macroscopic spatial clustering characteristics, thereby compensating for the limitations of RDR experiments that focus only on microscopic viewpoint pair relations. This experiment evaluates the extent to which encryption algorithms change spatial clustering characteristics and the accurate recovery ability of decryption algorithms by comparing the number, distribution and density changes of hotspots in the original, encrypted and decrypted data. Considering the requirement of privacy protection, an effective encryption algorithm should be able to alter the spatial clustering characteristics of the encrypted data and cause significant changes in the hotspot distribution pattern. Meanwhile, the decryption algorithm should completely recover the original hotspot characteristics to ensure the accuracy of data analysis.

1) Experimental design of hotspot area recognition

The experiments employ the density-based spatial clustering of applications with noise (DBSCAN) density clustering algorithm to identify hotspot areas in the data. The DBSCAN algorithm is chosen for its adaptability to clustering of irregular shapes and its inherent capability to handle noisy data. The algorithm controls the clustering process through two critical parameters, i.e., the neighborhood radius eps and the minimum number of points min_samples. For the original and decrypted data, the distance is calculated using the Haversine formula. In contrast, Euclidean distances are employed for encrypted data, as the coordinates of encrypted data have lost their geographical significance.

The neighborhood radius parameter is determined according to the data type. The eps for the original and decrypted data is set to 0.005 degrees, while the encrypted data is adaptively calculated according to its coordinate range. The optimized parameter selection ensures consistent clustering performance while adapting to the distribution characteristics of different data types.

TABLE II
STATISTICAL PROPERTIES OF THE SPATIAL DISTRIBUTION OF THE THREE TYPES OF DATA

| Data types | Longitude range | Latitude range | Number of outliers |
|---|---|---|---|
| Original data | 0-168.424730 | 0-50.375760 | 22,570 |
| Encrypted data | 0.8-179.959330 | 0.8-83.991250 | 21,026 |
| Decrypted data | 0-168.424730 | 0-50.375760 | 22,570 |

To ensure accuracy of the experimental results, the data are cleaned before sampling to remove records containing invalid coordinates. In order to avoid the interference of outliers on clustering a consistent sampling method is designed to extract the corresponding points in the three datasets. The method first identifies the trajectory points that exist in the original, encrypted and decrypted data at the same time, and then obtains representative 95,551 sample points by stratified random sampling. The sampling process stratifies the vehicle IDs to maintain trajectory diversity, while random seed fixation is used to ensure reproducible results.

To ensure accuracy of the experimental results, the data are cleaned before sampling to remove records containing invalid coordinates. In order to avoid the interference of outliers on clustering a consistent sampling method is designed to extract the corresponding points in the three datasets. The method first identifies the trajectory points that exist in the original, encrypted and decrypted data at the same time, and then obtains representative 95,551 sample points by stratified random sampling. The sampling process stratifies the vehicle IDs to maintain trajectory diversity, while random seed fixation is used to ensure reproducible results.

2) Results of the experiment

The experiments are conducted to analyze the spatial distribution of the three types of data. The statistical results presented in Table II show that the encryption process significantly changes the spatial extent of the data. This reconfiguration of coordinate space fundamentally destroys the original spatial distribution pattern, making it difficult for external observers to infer sensitive information through spatial distribution features.

The distribution properties of the decrypted data are identical to the original data, indicating that the decryption algorithm could accurately recover the original spatial structure. This result complements the RDR experiment, demonstrating that the encryption algorithm changes the microscopic viewpoint pair relations. While this experiment presents that the decryption algorithm restores the macroscopic spatial structure.

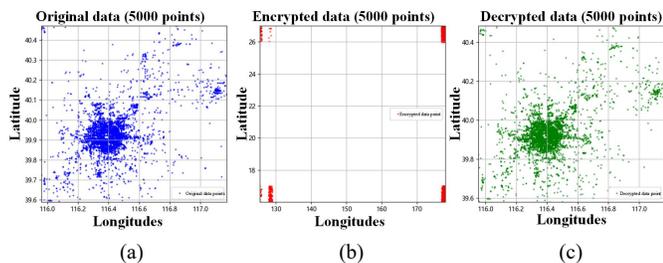

(a)          (b)          (c)

**Fig. 6.** Map of the spatial distribution of the three types of data

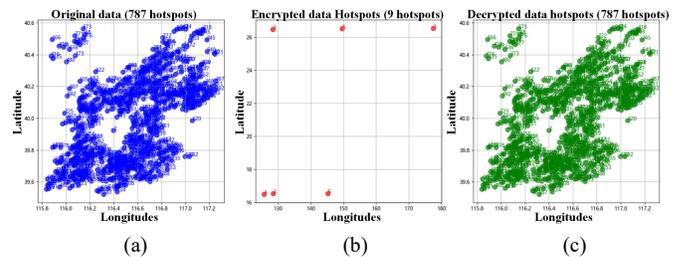

(a)          (b)          (c)

**Fig. 8.** Map of hotspot distribution for the three types of data

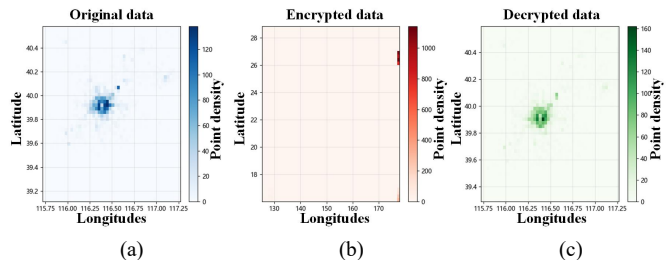

(a)          (b)          (c)

**Fig. 7.** Density profiles of the three types of data

The spatial distribution of the three types of data is analyzed in this experiment. As shown in Figure 6(a), the original data show a clear aggregation pattern in coordinate space. The points are mainly distributed with longitude ranging from 116.0 to 117.2 and latitude ranging from 39.6 to 40.5. As shown in Figure 6(b), the spatial distribution of the encrypted data changes drastically. The points are mainly distributed in two regions around the longitude at 130 and 170. The latitude ranges from 16.5 to 26.5. As shown in Figure 6(c), the distribution pattern of the decrypted data is almost identical to that of the original data. This indicates that the decryption algorithm has successfully recovered the original spatial structure.

This paper further demonstrates the spatial aggregation characteristics of the three types of data through density heat maps. As shown in Figure 7(a), the original data forms a clear high-density region in coordinate space, and the core area with the highest density is located near the longitude 116.5 as well as the latitude 39.9. As shown in Figure 7(b), the density distribution of the encrypted data exhibits significant variation, resulting in highly dispersed regions with density values nearly 10 times greater than those of the original data. As shown in Figure 7(c), the density distribution of the decrypted data demonstrates a high consistent with that of the original data, maintain both the same location of the density center and the same change in the density gradient.

The clustering results of the DBSCAN algorithm on the three types of data reveal the profound effect of encryption on the spatial clustering properties. As shown in Figure 8(a), 787 hotspots are identified in the original data, and these hotspots exhibit complex spatial distribution patterns. As shown in Figure 8(b), only nine hotspots are identified in the encrypted data, and the distribution pattern is completely different from the original data. As shown in Figure 8(c), the number of hotspots identified in the decrypted data is exactly the same as that of the original data, and the distribution patterns are also highly similar. The number of hotspots in the critical area decreased by 98.9% after encryption.

The drastic change in the number of hotspots proves that the encryption algorithm successfully destroys the spatial clustering properties of the original data, fundamentally changing the hotspot distribution pattern. In contrast, the perfect recovery of hotspots in the decrypted data, which is 100% number matching rate, indicates that the decryption algorithm could accurately restore the original clustering characteristics. The location matching analysis of the hotspot of both the original and decrypted data presents that the locations of 787 pairs of hotspots completely overlaps, with an average distance of 0 km and a location matching accuracy of 100%. This perfect matching degree not only shows the accuracy of the decryption algorithm in restoring the hotspot locations but also proves that it could completely recover the shape and boundary of the hotspots.

*C. Experimental Design of Decryption Accuracy*

The decryption accuracy test is an important part of evaluating the effectiveness of the decryption algorithm for automotive geographic data. This section quantitatively evaluates the precise recovery ability of the decryption algorithm by comparing the consistency between the original data and the decrypted data. The ideal decryption algorithm should perfectly restore the original data characteristics, thereby ensuring the availability of data and the accuracy of analysis results under authorized conditions.

1)  Experimental principles of decryption accuracy

In this experiment, the decryption accuracy is evaluated using the point-to-point exact matching method. The T-drive cab track dataset is used in this experiment, which contains a large number of real-world location records. This study compares the latitude and longitude coordinates of the original and the corresponding decrypted locations, and requires the coordinates to be exactly equal before the matching is considered successful. This strict standard could detect most accuracy deviation and provide a harsh reference benchmark for algorithm evaluation.

A metric method to accurately quantify the decryption accuracy is experimentally designed. Overall matching rate (OMR) is defined as the proportion of identical points to the overall number of points, and the file-level matching rate (FMR) is defined as the proportion of identical points in each file to the total number of points in that file. Besides, the mismatch rate (MMR) is defined as the proportion of imperfectly identical points to the overall number of points.

TABLE III
DECRYPTION ACCURACY STATISTICS RESULTS

| Index of correlation | Numerical value |
| --- | --- |
| The total number of location points | 17,374,292 |
| Exactly the same number of points | 17,374,292 |
| Different points | 0 |
| Overall matching rate | 100% |
| Mismatch rate | 0% |
| The number of processed files | 10,357 |
| The exact number of matching files | 10,357 |

The experiment is conducted with strict point-to-point comparison. This metric method constructs file correspondences, accurately compares each pair of latitude and longitude coordinates, and ultimately calculates key indicators while presenting the results through visualization.

2) Results of the experiment

This study conducts a comprehensive decryption accuracy on the T-drive dataset, and the experimental results are shown in Table III. The overall matching rate reaches a rare 100%, and all 17,374,292 position points are fully consistent with the original data after decryption, and no mismatched points are found. In terms of file level, all 10,357 trajectory files reach 100% matching rate, indicating that the decryption algorithm exhibits a highly consistent reduction capability when dealing with different vehicle trajectory data.

The experimental results show that the decryption algorithm achieves completely accurate data reduction on all test data. 100% matching rate proves the reliability and accuracy of the proposed decryption method. Considering the data security, this perfect matching means that the encrypted data could be completely restored to its original form under authorized circumstances without any loss of accuracy of vehicle location data due to the decryption process.

## V. CONCLUSION

This paper proposes a high-precision privacy protection mechanism based on FPE of automotive geographical coordinates. Through composite key design, improved Feistel network and geographic range constraints, the mechanism achieves encrypted data format compatibility, spatial relationship perturbation and zero decryption error, providing an effective solution for automotive location data security in intelligent transportation systems.

The effectiveness of the suggested mechanism is confirmed by experiments conducted on the T-drive dataset. The RDR results show that spatial relationship perturbation is effective for robust location privacy protection. After encryption, the number of hotspots in the critical area decreased by 98.9%, indicating that spatial clustering characteristics were successfully destroyed. The decryption algorithm guarantees 100% accuracy in data analysis across over 17 million location points under authorized conditions.

Although our method has achieved remarkable results, there are still several limitations in this work. Specifically, the method relies on offline data validation and its inadequate defense against sophisticated hybrid attacks. To ensure the security for multimodal spatiotemporal data, future research will concentrate on 3D coordinate encryption, dynamic parameter optimization, and differential privacy enhancement.

ACKNOWLEDGMENT

This work was supported in part by the National Natural Science Foundation of China under Grant 52102447, in part by the R&D Program of Beijing Municipal Education Commission under Grant KM202411232005, and in part by the 2023 Industrial Internet Enterprise Data Security Joint Management and Control System Project under Grant 0747-2361SCCZA195.

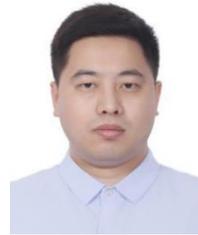

**Haojie Ji** received the M.S. degree from the School of Mechanical Engineering, Yanshan University, Qinhuangdao, China, in 2015, and the Ph.D. degree from the School of Transportation Science and Engineering, Beihang University, Beijing, China, in 2019. He is currently an Associate Professor with the College of Mechanical and Electrical Engineering, Beijing Information Science & Technology University, Beijing, China. His research interests include cyber security and privacy protection in autonomous driving.

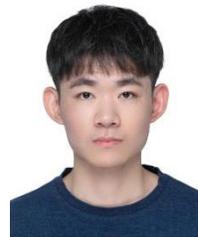

**Long Jin** received the B.E. degree from Tiangong University, Tianjin, China, in 2022. He is currently pursuing the master's degree with the College of Mechanical and Electrical Engineering, Beijing Information Science & Technology University, Beijing, China, under the supervision of Associate Professor Haojie Ji. His research interests include data security risk assessment and multimodal data privacy protection in autonomous driving.

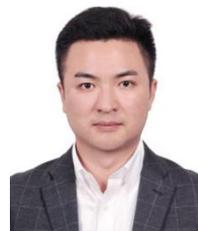

**Haowen Li** (Member, IEEE) received the B.S. degree from China University of Petroleum, Beijing, China, in 2010, and the M.S. degree from the Hong Kong Polytechnic University, Hong Kong, China, in 2024. He is currently pursuing the Ph.D. degree with the College of Faculty of Engineering, Hong Kong Polytechnic University, Hong Kong, China. His research interests include cyber security, privacy, decentralized identity, and IoT security, especially intelligent connected vehicles.

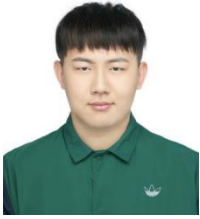
**Chongshi Xin** received the B.E. degree from North China Institute of Aerospace Engineering , China, in 2021. He is currently pursuing the master's degree with the College of Mechanical and Electrical Engineering, Beijing Information Science & Technology University, Beijing, China, under the supervision of Professor Liyong Wang. His research interests include data security privacy protection in autonomous driving.

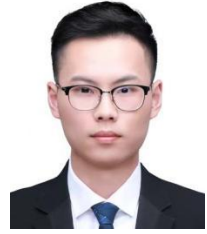
**Te Hu** received the B.E. degree in Engineering from China University of Mining and Technology, Xuzhou, China, in 2020. He is currently pursuing the master's degree with the College of Mechanical and Electrical Engineering, Beijing Information Science & Technology University, Beijing, China, under the supervision of Associate Professor Haojie Ji. His research interests include intelligent driving security, adversarial attacks and defenses in autonomous systems.